\documentclass[12pt,a4paper]{article}
\usepackage{amsfonts,latexsym}
\usepackage{amsmath,amssymb}
\usepackage{graphicx,color}
\usepackage{multirow}
\usepackage{slashbox}
\numberwithin{equation}{section} \oddsidemargin 0 mm \evensidemargin
0 mm \topmargin -10 mm \textheight 215 mm \textwidth 163 mm

\renewcommand{\thefootnote}{\fnsymbol{footnote}}
\newcommand{\nn}{\nonumber}

\begin{document}
\vspace{12mm}

\begin{center}
{{{\Large {\bf Polarization modes of gravitational waves \\ in three-dimensional massive gravities}}}}\\[10mm]

{Taeyoon Moon$^{a}$\footnote{e-mail address: tymoon@sogang.ac.kr}
and  Yun Soo Myung$^{b}$\footnote{e-mail address:
ysmyung@inje.ac.kr},
}\\[8mm]

{{${}^{a}$ Center for Quantum Space-time, Sogang University, Seoul, 121-742, Korea\\[0pt]
${}^{b}$ Institute of Basic Sciences and School of Computer Aided Science, Inje University Gimhae 621-749, Korea}\\[0pt]
}
\end{center}
\vspace{2mm}

\begin{abstract}
We find polarization modes of gravitational waves in  topologically
massive and new massive gravities by using the Newman-Penrose
formalism where the null real tetrad is necessary to specify
gravitational waves. The number of polarization modes is two for the
new massive gravity and  one for the topologically massive gravity,
which is consistent with the metric-perturbation approach.
\end{abstract}

{\footnotesize ~~~~PACS numbers: }

\vspace{1.5cm}

\hspace{11.5cm}{Typeset Using \LaTeX}
\newpage
\renewcommand{\thefootnote}{\arabic{footnote}}
\setcounter{footnote}{0}


\section{Introduction}
Einstein gravity  has been known to have  no propagating degrees of
freedom in three dimensions.   Massive generalizations of the
Einstein gravity may allow propagating degrees of freedom.
Topologically massive gravity (TMG) is the  famous gravity theory
obtained by including  a gravitational Chern--Simons term (gCS) with
coupling $\mu$~\cite{DJT, DJT2}. Since the gCS term is odd under
parity, the theory shows  a single massive propagating degree of
freedom of a given helicity, whereas the other helicity mode remains
massless.   The model was extended by adding  a cosmological
constant $\Lambda=-1/\ell^2$  to the topologically massive
gravity~\cite{Deser82}.  Then, the single massive field could be
realized as a massive scalar $\varphi=z^{3/2}h_{zz}$ when employing
the Poincare coordinates $x^{\pm}$ and $z$  covering the AdS$_3$
spacetimes~\cite{CDWW}.  It was shown that the massive graviton
having negative-energy disappears at the chiral point of $\mu\ell=1$
by Lee-Song-Strominger in Ref.~\cite{LSS}. Furthermore, this
cosmological topological massive gravity at the chiral point may be
described by the logarithmic conformal field theory
~\cite{GJ,Myung}. Importantly,   the Lee-Song-Strominger work has
indicated that the ``third-order" Einstein equation  turned out to
be the ``first-order" equation for a massive graviton when choosing
the transverse-traceless gauge for metric tensor.

On the other hand,  Bergshoeff, Hohm, and Townsend have recently
proposed another massive generalization of the Einstein gravity by
adding a specific quadratic curvature term   to the Einstein-Hilbert
action~\cite{bht,bht2}.  This term was designed to reproduce  the
ghost-free Fierz-Pauli action for a massive propagating graviton in
the linearized approximation, whereas it differs from the
Fierz-Pauli term when considering the non-linear terms. This gravity
theory became known as new massive gravity (NMG). Unlike the TMG,
the NMG preserves parity.  As a result, the gravitons acquire the
same mass for both helicity states, indicating two massive
propagating degrees of freedom. Considering TMG together with NMG
leads to mass-splitting between helicity states.

So far, we have considered only the conventional metric-perturbation
approach to three-dimensional massive gravities. Hence, we do not
know explicitly what are polarization modes of gravitational waves
(GW) in  TMG and NMG. Since two massive theories belong to higher
curvature gravity, we need to introduce the Newman-Penrose formalism
\cite{Newman} where the null real tetrad is necessary to specify
polarization modes of GW, as the four-dimensional massive gravity
requires  null complex tetrad  to specify six independent
polarization modes of $\{ \Phi_2, \Psi_3, \Psi_4, \Phi_{22}\}$
\cite{Eardley:1974nw}. Here $\Psi_3$ and $\Psi_4$ are complex, and
analyzing the rotational behavior of the set shows the respective
helicity values $s=\{0,\pm1,\pm2,0\}$. It was suggested that the
observations of the GW will be done in the near future, and the
corresponding determination of all possible states of polarization
would be a very powerful test to rule out the present studied
alternative theories of gravity.

In this work, we will find the polarization modes of gravitational
waves arisen from the TMG and NMG by employing  the the
Newman-Penrose formalism in three dimensions. Even though these
theories are not four-dimensional gravity theory, they will provide
a prototype of polarization states.   This work will be important to
see how massive modes with different polarizations propagate in the
three-dimensional Minkowski spacetimes. Since higher-order Einstein
equation becomes lower-order equation when using the linearized
Ricci tensor $R^L_{\mu\nu}$  instead of metric tensor $h_{\mu\nu}$,
this work will provide another approach  in addition to the
conventional metric-perturbation theory.

\section{Null real triad formalism  in three dimensions}
Let us first introduce  a triad\footnote{Note that in \cite{Hall}
they have used  the different notation with metric signature
$(+,+,-)$. This work shall use the notations employed in
\cite{Sousa:2007ax}. A triad may be  defined by using complex basis.
However, in this case,  we could not describe a propagating mode
because it provides  the stationary wave only.}
\cite{Hall,Sousa:2007ax} of real vectors $\{ {\bf
k},\bf{n},\bf{m}\}$ which are related to the Cartesian tetrad
vectors $\{{\bf e}_{t},{\bf e}_{x},{\bf e}_{z}\}$ in  three
dimensions with metric signature $(-,+,+)$ as
\begin{equation}
{\bf k}=\frac{1}{\sqrt{2}}({\bf e}_t+{\bf e}_z),~~ {\bf
n}=\frac{1}{\sqrt{2}}({\bf e}_t-{\bf e}_z),~~ {\bf m}={\bf e}_x,
\end{equation}
where they  satisfy the relations
\begin{eqnarray}
-{\bf k}\cdot{\bf n}={\bf m}^2=1,~~~ {\bf k}\cdot{\bf m}={\bf
n}\cdot{\bf m}={\bf k}^2={\bf n}^2=0.
\end{eqnarray}
Note that a tensor ${\bf T}$  can be written as
\begin{eqnarray}\label{tensor}
T_{abc...}=T_{\mu\nu\rho...}a^{\mu}b^{\nu}c^{\rho_{...}},
\end{eqnarray}
where $(a,b,c,...)$ run over  $( {\bf k},\bf{n},\bf{m})$ and
$(\mu,\nu,\rho,...)$ run over $(t,~x,~z)$.
 It is well-known that the
Weyl tensor vanishes identically in three dimensions.  Therefore,
the Riemann tensor with six independent components can be decomposed
into the Ricci tensor and Ricci scalar as
\begin{eqnarray}
R_{\rho\sigma\mu\nu}=2\Big(g_{\rho[\mu}R_{\nu]\sigma}
-g_{\sigma[\mu}R_{\nu]\rho}\Big)-Rg_{\rho[\mu}g_{\nu]\sigma}.
\end{eqnarray}
On the other hand,  by using the formalism of real two-component
spinors \cite{Sousa:2007ax}, the Ricci spinor $\Phi_{ABCD}$ can be
expressed in terms of the Ricci tensor as
\begin{eqnarray}
\Phi_{00}&\equiv&\Phi_{0000}=\frac{1}{2}R_{\mu\nu}k^{\mu}k^{\nu},\nn\\
\Phi_{22}&\equiv&\Phi_{1111}=\frac{1}{2}R_{\mu\nu}n^{\mu}n^{\nu},\nn\\
\Phi_{10}&\equiv&\Phi_{1000}=\frac{1}{2\sqrt{2}}R_{\mu\nu}m^{\mu}k^{\nu},\nn\\
\Phi_{12}&\equiv&\Phi_{1011}=\frac{1}{2\sqrt{2}}R_{\mu\nu}m^{\mu}n^{\nu},\nn\\
\Phi_{11}&\equiv&\Phi_{0011}=
\frac{1}{6}\left(R_{\mu\nu}m^{\mu}m^{\nu}+R_{\mu\nu}n^{\mu}k^{\nu}\right).
\label{Phi}
\end{eqnarray}
 Eardley et al.\cite{Eardley:1974nw} have shown that
polarization states of the GW in four dimensions can be given by six
independent components of the Riemann tensor.  They assumed that the
GW are weak and take nearly plane waves propagating in the $+z$
direction.  Accordingly,  the GW have  six independent modes which
correspond to six independent Riemann tensor of the Newman-Penrose
tetrad.   Following the Eardley et al. approach, we  consider the
plane GW propagating in the $+z$ direction, which means that all
quantities have the forms of $(t-z)$ only. This is equivalent to
gauge-fixing (for example, transverse gauge) in the
metric-perturbation approach. In this case,  it is shown  that the
Riemann tensor satisfies the following relation
\begin{eqnarray}\label{iden}
\partial_{p}R_{abcd}=0,
\end{eqnarray}
where $(a,b,c,d$) run over  in $( {\bf k},\bf{n},\bf{m})$ and
$(p,q)$ run over $(\bf{k},{\bf m})$.  Introducing a metric
perturbation of $g_{\mu\nu}=\eta_{\mu\nu}+h_{\mu\nu}$ and the
linearized Riemann tensor $R_{abcd}^{L}(h)$, the Bianchi identity
can be written as
\begin{eqnarray}\label{iden2}
\nabla_{[n} R_{pq]ab}=0\Rightarrow\partial_{[n}
R_{pq]ab}^{L}&=&\frac{1}{3}\left(\partial_{n} R_{pqab}^{L}
+\partial_{p} R_{qnab}^{L}+\partial_{q} R_{npab}^{L}\right)\nn\\
 &=&\frac{1}{3}\partial_{n} R_{pqab}^{L}=0.
\end{eqnarray}
In deriving the second line, we have used the relation (\ref{iden}).
Consequently, from  Eq. (\ref{iden2}), we have
\begin{equation}
R_{pqab}^{L}(h)={\rm const.},\end{equation}  which implies that the
independent  components of the Riemann tensor are given by three  of
$R_{pnqn}^{L}$  as
\begin{eqnarray}\label{3riemann}
R_{knkn}^{L}(h),~~~R_{knmn}^{L}(h),~~~R_{mnmn}^{L}(h).
\end{eqnarray}
It is noted  that three components (\ref{3riemann})  can be
reexpressed  in terms of the Ricci tensor\footnote{In $\{{\bf
k,n,m}\}$ basis, the Ricci tensor is given by
$R_{ac}=g^{bd}R_{abcd}$ with $g^{bd}=\left(
\begin{array}{ccc}
0 & -1 & 0 \cr -1 & 0 & 0 \cr 0 & 0 & 1
\end{array}
\right)$.} in $\{{\bf k,n,m}\}$ basis like
\begin{eqnarray}
R_{kn}^{L}(h),~~~~~R_{mn}^{L}(h),~~~~~R_{nn}^{L}(h)
\end{eqnarray}
which correspond to $\Phi_{11},~\Phi_{12},$ and $\Phi_{22}$ in
(\ref{Phi}), respectively.

Finally, we again  point out that in three dimensions,  the maximum
number of polarization modes  for GW  is three of
$R_{kn}^{L},~R_{mn}^{L}~,$ and $R_{nn}^{L}$ (see Figure
1)\footnote{In order to show  the polarization modes of weak, plane
GW explicitly, we first consider the geodesic deviation equation (or
relative accelerations between nearby particles) as was shown in
\cite{Eardley:1974nw}: $\frac{\partial^2}{\partial
t^2}S^{\mu}=a^{\mu}=S^{\sigma}R^{\mu}_{~tt\sigma}$, where $S^{\mu}$
is a vector field measuring the deviation between geodesics. Using
the geodesic deviation equation, we  check easily that in three
dimensions the Riemann tensor has  three components of
$R^{x}_{~ttx}$, $R^{x}_{~ttz}$, and $R^{z}_{~ttz}$. Accordingly, we
may  draw $S^{\mu}$ for three modes by considering the geodesic
deviation equation and $R^{\mu}_{~tt\sigma}=R^{\mu}_{~tt\sigma}(u)$
where $u=t-z$, as depicted in Fig.1.}. In the next section we will
investigate these modes by considering three gravity theories.

\section{Massive gravities}
We mention that the procedure of determining the number of the
independent component  of  the Riemann tensor was done by choosing
the plane wave solution to the vacuum  linearized Einstein equation.
This implies that the observer is far from GW sources, which means
that it is enough to solve the vacuum linearized Einstein equation.
In order to find polarization modes, we introduce three theories of
the Einstein gravity, NMG and TMG. Before finding polarization modes
from three theories,  we wish to mention  the  three dimensional
Fierz-Pauli (FP) massive  equations which will be used  to define
the spin 2 in Sec.3.3 and  3.4.

\subsection{FP massive  equations}
It is well known that the  FP massive equations for a symmetric
rank-$s$ tensor field describe the massive modes of helicity $\pm s$
with mass $m$.  In three-dimensional Minkowski spacetimes, the FP
massive equations  are given by~\cite{Berg}
\begin{eqnarray}\label{fp1}
[{\cal D}(m){\cal
D}(-m)]_{~\mu_1}^{\rho}\phi_{\rho\mu_2\cdots\mu_s}=0,
~~~~\eta^{\mu\nu}\phi_{\mu\nu\rho_1\cdots\rho_{s-2}}=0.
\end{eqnarray}
Here the operator
\begin{eqnarray}
{\cal D}(m)_{\mu}^{\nu} =\frac{1}{2}\Big[\delta_{\mu}^{\nu}
-\frac{1}{m}\epsilon_{\mu}^{~\rho\nu}\partial_{\rho}\Big]
\end{eqnarray}
is an on-shell projection operator as
\begin{equation}
{\cal D}^2(m) \phi={\cal D}(m)\phi
\end{equation}
if $\phi$ satisfies (\ref{fp1}). In particular, if one considers the
generalized massive gravity of NMG + gCS term,  the parity-violating
FP equations takes the form
\begin{eqnarray}
[{\cal D}(m_+){\cal
D}(-m_{-})]_{~\mu_1}^{\rho}\phi_{\rho\mu_2\cdots\mu_s}=0,
~~~~\eta^{\mu\nu}\phi_{\mu\nu\rho_1\cdots\rho_{s-2}}=0
\end{eqnarray}
for two independent masses $m_\pm$.   These equations show that one
mode of helicity $s$ with mass $m_+$ and the other of helicity $-s$
with mass $m_{-}$ propagate. In this case,  the second-order
dynamical equation leads to
\begin{eqnarray} \label{sodeq}
\left(\Box-m^2\right)\phi_{\mu_1\cdots\mu_s} =\tilde{\mu}
\epsilon_{\mu_1}^{~~\rho\nu}\partial_{\rho}\phi_{\nu\mu_2\cdots\mu_s}
\end{eqnarray}
with \begin{equation} m^2=m_+m_{-},~~\tilde\mu\equiv
m_{-}-m_{+}.\end{equation}
 In the  limit of
$m_{-}\to\infty$ for fixed $m_{+}$, the helicity $-s$  mode
decouples and thus, a single mode of helicity $s$ and  mass $m_+$ is
described by the first-order equation
\begin{eqnarray}\label{fp3}
{\cal D}(\mu)_{~\mu_1}^{\rho}\phi_{\rho\mu_2\cdots\mu_s}=0,
~~~~\eta^{\mu\nu}\phi_{\mu\nu\rho_1\cdots\rho_{s-2}}=0,
\end{eqnarray}
where $\mu=m^2/\tilde{\mu}$. Let us confine ourselves to spin 2, and
consider the self-dual spin 2 model with field equation
\begin{equation} \label{eqq1}
{\cal
D}(\mu)^\rho_{~\mu}\phi_{\rho\nu}=0,~~\eta^{\mu\nu}\phi_{\mu\nu}=0
\end{equation}
and the subsidiary condition $\partial^\mu \phi_{\mu\nu}=0$.  The
general solution is given by
\begin{equation}
\phi_{\mu\nu}={\cal G}^{\rho\sigma}_{\mu\nu}h_{\rho\sigma} \equiv
G^{\rm L}_{\mu\nu},~~{\cal
G}^{\rho\sigma}_{\mu\nu}=-\frac{1}{2}\epsilon_{(\mu}^{~~\eta
\rho}\epsilon_{\nu)}^{~~\tau\sigma}\partial_\eta\partial_\tau
\end{equation}
for some second-rank tensor $h_{\mu\nu}$. Here $G^L_{\mu\nu}$ is the
linearized Einstein tensor for the metric perturbation of
$g_{\mu\nu}=\eta_{\mu\nu}+h_{\mu\nu}$. Then, the self-dual field
equation of  (\ref{eqq1}) becomes
\begin{equation} \label{litmg}
{\cal
D}(\mu)^\rho_{~\mu}G^L_{\rho\nu}=0,~~\eta^{\mu\nu}G^L_{\mu\nu}=0
\end{equation}
which implies  the linearized Einstein equation for the TMG. This
confirms the equivalence of the linearized TMG to the self-dual spin
2 theory in three dimensions.

 Similarly, we check that  the FP
massive equation
\begin{equation} \label{fpeq}
[{\cal D}(m){\cal D}(-m)]_{~\mu}^{\rho}\phi_{\rho\nu}=0,
~~~~\eta^{\mu\nu}\phi_{\mu\nu}=0
\end{equation}
is equivalent to the linearized equations for the NMG
\begin{equation}
[{\cal D}(m){\cal D}(-m)]_{~\mu}^{\rho}G^L_{\rho\nu}=0, ~~R^L=0.
\end{equation}

\begin{figure*}[t!]
   \centering
   \includegraphics{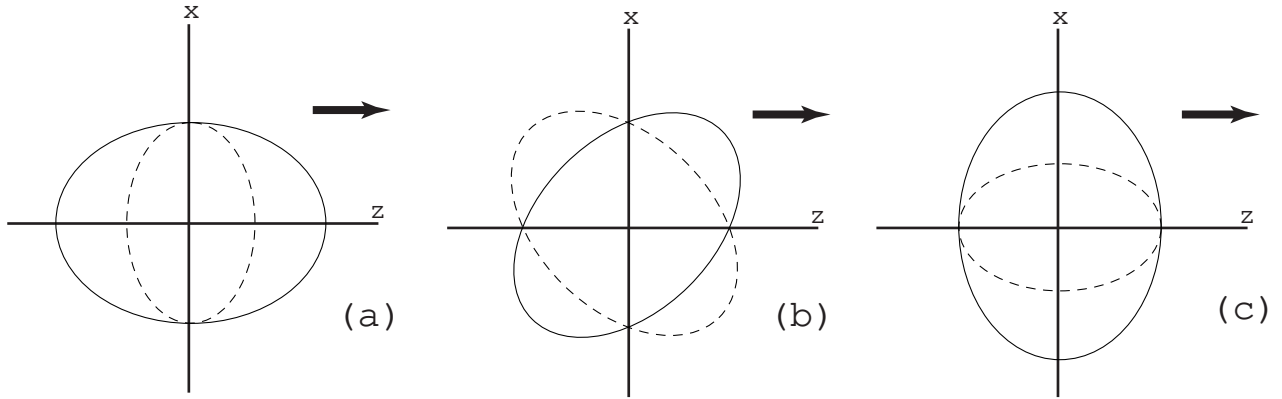}
\caption{Three polarization modes of weak, plane GW permitted in
three-dimensional massive gravity. The displacement shows that each
mode induces on a sphere of test particles, i.e., $(a):~
\Phi_{11}=\frac{1}{6}R_{kn}=\frac{1}{6}R_{ztzt},
~(b):~\Phi_{12}=\frac{1}{2\sqrt{2}}R_{mn}=\frac{1}{2}R_{xtzt},
~(c):~\Phi_{22}=~\frac{1}{2}R_{nn}=R_{xtxt}$. In this figure all
waves are propagating in the $+z$ direction. } \label{Rnn}
\end{figure*}
\subsection{Einstein gravity}
The Einstein-Hilbert action with a matter term is given by
\begin{eqnarray}
S_{EH}=\frac{1}{16\pi G}\int d^{3}x \sqrt{-g}R+S_m,
\end{eqnarray}
which yields the Einstein equation \begin{equation} G_{\mu\nu}=8\pi
G
T_{\mu\nu},~~~G_{\mu\nu}=R_{\mu\nu}-\frac{R}{2}g_{\mu\nu}.\end{equation}
In the case of $T_{\mu\nu}=0$, we obtain $R_{\mu\nu}=0$. Considering
the metric perturbation of $g_{\mu\nu}=\eta_{\mu\nu}+h_{\mu\nu}$,
the linearized  equation becomes
\begin{eqnarray}
R_{\mu\nu}^{L}(h)=0.
\end{eqnarray}
Using the relation (\ref{tensor}) between $(\mu,\nu,\rho,...)$ and
$(a,b,c,...)$,  one shows that $R_{\mu\nu}^{L}=0$ correspond to
\begin{eqnarray}
R_{kn}^{L}=R_{mn}^{L}=R_{nn}^{L}=0 \to
\Phi_{11}=\Phi_{12}=\Phi_{22}=0.
\end{eqnarray}
This indicates that there is no propagating mode of the GW in
Einstein gravity.  However, the result is nothing  new because  the
linearized second-order equation for $h_{\mu\nu}$ implies no
graviton in three dimensions~\cite{Deser:1983tn}. We would like to
mention that the Newman-Penrose approach confirms the result of the
metric-perturbation approach.

\subsection{NMG}
In NMG \cite{Bergshoeff:2009hq} proposed by Bergshoeff, Hohm, and
Townsend,  the action is given by
\begin{eqnarray}\label{actionnmg}
S_{\rm NMG}=\frac{1}{\kappa^2}\int d^{3}x\sqrt{-g}\Big(-R+\alpha
R^2+\beta R_{\mu\nu}R^{\mu\nu}\Big),
\end{eqnarray}
where the constant $\kappa$ has the mass dimension $[\kappa]=-1/2$,
and $\alpha $ and $\beta$ are dimensionless constants satisfying the
important relation of  $\alpha=-3\beta/8$ which kills the spin-0
mode (scalar graviton). The wrong sign in the Einstein-Hilbert term
is necessary to avoid the ghost. From the action (\ref{actionnmg}),
the equation of motion for the metric can be derived as
\begin{eqnarray}\label{nmgeq}
&&G_{\mu\nu}-\alpha\Big[2RG_{\mu\nu}+2g_{\mu\nu}\nabla_{\gamma}\nabla^{\gamma}R
-2\nabla_{\mu}\nabla_{\nu}R\Big]\nn\\
&&\hspace*{-1em}-\beta\Big[-\frac{1}{2}g_{\mu\nu}R_{\rho\sigma}R^{\rho\sigma}
+2R_{\mu\rho\nu\sigma}R^{\rho\sigma}+\nabla_{\gamma}\nabla^{\gamma}
R_{\mu\nu}+\frac{1}{2}g_{\mu\nu}\nabla_{\gamma}\nabla^{\gamma}R
-\nabla_{\mu}\nabla_{\nu}R\Big]=0.
\end{eqnarray}
Considering the metric perturbation
$g_{\mu\nu}=\eta_{\mu\nu}+h_{\mu\nu}$, the linearized equation takes
the form
\begin{eqnarray}\label{nmgeq2}
 \square
 R_{\mu\nu}^{L}-\frac{1}{\beta}R_{\mu\nu}^{L}=0,~~\square=\partial_\gamma\partial^\gamma
\end{eqnarray}
which is interpreted as   the second-order equation for
$R^L_{\mu\nu}$.  This means that the fourth-order equation for
$h_{\mu\nu}$ could be interpreted as the second-order equation for
$R^L_{\mu\nu}(h)=-\frac{1}{2}\square h^{TT}_{\mu\nu}$ with
transverse-traceless (TT) gauge. In deriving this, we have used the
perturbation equation for the trace of Eq.(\ref{nmgeq}) like
\begin{eqnarray}\label{nmgR}
R^{L}=0
\end{eqnarray}
together with $8\alpha+3\beta=0$. It is important to note  that
Eq.(\ref{nmgeq2}) is exactly the same with the second-order
dynamical equation (\ref{sodeq}) with $m_+=m_-=m$ when replacing
$\phi_{\mu\nu}\Leftrightarrow R_{\mu\nu}^{L}$ and
$m\Leftrightarrow\frac{1}{\sqrt{\beta}}$. This implies that
Eq.(\ref{nmgeq2}) describes the massive modes of helicity $\pm 2$
with mass $1/\sqrt{\beta}$.

Furthermore, the solution to Eq.(\ref{nmgeq2}) is given by
\begin{eqnarray}\label{nmgRmunu}
R_{\mu\nu}^{L}=M_{\mu\nu}e^{ip\cdot x},
\end{eqnarray}
where $p^2=-1/\beta$ and $M_{\mu\nu}$ is a constant symmetric tensor
which satisfies to the following conditions:
\begin{eqnarray}\label{TTc}
M^{\mu}_{\mu}=0,~~~p^{\mu}M_{\mu\nu}=0.
\end{eqnarray}
 From (\ref{tensor}) and
the Ricci scalar in $\{{\bf k,n,m}\}$ basis, we find that
Eqs.(\ref{nmgR}) and (\ref{nmgRmunu}) correspond to
\begin{eqnarray}
R_{kn}^{L}=0~~~{\rm and}~~~R_{mn}^{L}\neq0,~~R_{nn}^{L}\neq0,
\end{eqnarray}
respectively. This shows that the number of  polarization modes of
GW in the NMG is  two of  $R_{mn}^{L}\to \Phi_{12}$ and
$R_{nn}^{L}\to \Phi_{22}$. Interestingly,  as was shown in
\cite{Bergshoeff:2009hq},  two propagating  modes appear in the NMG.
It is worth noting  that in our approach, $R_{kn}^{L}=0$ implies
that there is no  ghost-like massive mode of zero helicity because
of $R_{kn}^{L}=R^{L}=0$. In addition, $R_{mn}^{L}$ and $R_{nn}^{L}$
correspond to the TT~\footnote{In our approach, we note that the
conditions (\ref{TTc}) correspond to the TT gauge in the
metric-perturbation theory.} metric-perturbation theory. Their two
polarization modes are depicted in $(b)$ and $(c)$ of Fig. 1.

 On the other hand, we may consider the general case of
$8\alpha+3\beta\neq0$. In this case, instead of Eqs.(\ref{nmgeq2})
and  (\ref{nmgR}),  we obtain
\begin{eqnarray}\label{nmgeq3}
R_{\mu\nu}^{L}-\beta\square R_{\mu\nu}^{L}+
(2\alpha+\beta)\partial_{\mu}\partial_{\nu}R^L+(2\alpha+\beta)\eta_{\mu\nu}
\square R^L&=&0,\\
(8\alpha+3\beta)\square R^{L}+R^{L}&=&0.
\end{eqnarray}
The solutions for the above equations are obtained by
\begin{eqnarray}\label{nnmgsol}
R^{L}=R_0e^{iq\cdot x},~~~R_{\mu\nu}^{L}=N_{\mu\nu}e^{i q\cdot x},
\end{eqnarray}
where $q^2=1/(8\alpha+3\beta)$ and $N_{\mu\nu}$ is a constant
symmetric tensor given by
\begin{eqnarray}
N_{\mu\nu}=\frac{8\alpha+3\beta}{4}\left(q_{\mu}q_{\nu}-
\frac{\eta_{\mu\nu}}{8\alpha+3\beta}\right)R_0.
\end{eqnarray}
We see that in $\{{\bf k,n,m}\}$ basis, the solution (\ref{nnmgsol})
corresponds to
\begin{eqnarray}
R_{kn}^{L}\neq0,~~~~~R_{mn}^{L}\neq0,~~~~~R_{nn}^{L}\neq0,
\end{eqnarray}
which mean that there exist three independent polarization modes
($\Phi_{11},\Phi_{12},\Phi_{22}$) in the  NMG with
$8\alpha+3\beta\neq0$. Their three polarization
 modes are depicted in $(a)$, $(b)$, and $(c)$ of Fig. 1.

\subsection{TMG}

The TMG was first proposed by Deser, Jackiw, and Templeton
\cite{Deser:1982vy,Deser:1981wh} in the aim of making the massive
gravity  theory in three dimensions. The TMG action takes the form
\begin{eqnarray}\label{tmgs}
S_{TMG}=-\frac{1}{\kappa^2}\int
d^{3}x\sqrt{-g}R-\frac{1}{\mu\kappa^2}S_{CS},
\end{eqnarray}
where $\mu$ is the gCS coupling  constant and $S_{CS}$ is the gCS
term,
\begin{eqnarray}
S_{CS}=\frac{1}{2}\int
d^{3}x\sqrt{-g}\epsilon^{\lambda\mu\nu}\Gamma_{\lambda\sigma}^{\rho}
\Big(\partial_{\mu}\Gamma_{\rho\nu}^{\sigma}+\frac{2}{3}
\Gamma_{\mu\tau}^{\sigma}\Gamma_{\nu\rho}^{\tau}\Big).
\end{eqnarray}
The wrong sign in the Einstein-Hilbert term is necessary to avoid
the ghost. Varying the action (\ref{tmgs}) with respect to the
metric yields
\begin{eqnarray}\label{tmgeq}
G_{\mu\nu}+\frac{1}{\mu}C_{\mu\nu}=0,
\end{eqnarray}
where $C_{\mu\nu}$ is the Cotton tensor defined  by
\begin{eqnarray}
C_{\mu\nu}\equiv\epsilon_{\mu}^{~\alpha\beta}\nabla_{\alpha}\left(R_{\beta\nu}
-\frac{1}{4}g_{\beta\nu}R\right).
\end{eqnarray}
One finds that taking the trace of Eq.(\ref{tmgeq}) leads to $R=0$.
Substituting $R=0$ into Eq.(\ref{tmgeq}), the linearized equation
can be written by
\begin{eqnarray}\label{tmgRmunu}
R_{\mu\nu}^{L}=-\frac{1}{\mu}\eta_{\mu\rho}\epsilon^{\rho\alpha\beta}
\partial_{\alpha}R_{\beta\nu}^{L}
\end{eqnarray}
which leads to (\ref{litmg}) when replacing $\mu$ by $-\mu$.
Importantly, we point out that Eq.(\ref{tmgRmunu}) is the
third-order equation for $h_{\mu\nu}$, whereas it is regarded as the
first-order equation for $R^L_{\mu\nu}=-\frac{1}{2}\square
h_{\mu\nu}^{TT}$.  It is interesting to note that
Eq.(\ref{tmgRmunu}) is equivalent to the self-dual model with
(\ref{eqq1})  when replacing $\phi_{\mu\nu}\Leftrightarrow
R_{\mu\nu}^{L}$ and $\mu\Leftrightarrow -\mu$. Apparently, from Eq.
(\ref{tmgRmunu}) and $R^{L}=0$, we expect that  the independent
components of the Riemann tensor is two of $R_{mn}^{L}$ and
$R_{nn}^{L}$, as was shown in the NMG. However, we have  to check
whether $R_{mn}^{L}$ and $R_{nn}^{L}$ are truly independent
components  even though we do not know the exact solution. To this
end, we note that there are mixing between components of the Ricci
tensor in Eq.(\ref{tmgRmunu}) because of the Livi-Civita tensor
$\epsilon^{\rho\alpha\beta}$. From the $(t,t),~(t,z),$ and $(z,z)$
components of Eq.(\ref{tmgRmunu}), we obtain  one first-order
equation (see Appendix for details):
\begin{eqnarray}
\partial_{x}\Big(R_{tt}^{L}+2R_{tz}^{L}+R_{zz}^{L}\Big)
-\mu\Big(R_{tt}^{L}+2R_{tz}^{L}+R_{zz}^{L}\Big)=(\partial_t+\partial_z)
\Big(R_{tx}^{L}+R_{xz}^{L}\Big),
\end{eqnarray}
which  implies that $R_{tt}^{L}+2R_{tz}^{L}+R_{zz}^{L}$ and
$R_{tx}^{L}+R_{xz}^{L}$ are not independent. Using the $\{{\bf
k,n,m}\}$ basis,  this indicates that $R_{mn}^{L}$ and $R_{nn}^{L}$
are not independent because
$R_{mn}^{L}=-\left(R_{tx}^{L}+R_{xz}^{L}\right)/\sqrt{2}$ and
$R_{nn}^{L}=\left(R_{tt}^{L}+2R_{tz}^{L}+R_{zz}^{L}\right)/2$.

Finally, we conclude that there exists one independent mode of GW in
the TMG. If one chooses the positive gCS coupling with $\mu>0$, its
mode is $R_{nn}^{L}=\Phi_{22}$, while for $\mu<0$,  its mode is
$R_{mn}^{L}=\Phi_{12}$ or vice versa.

\section{Discussions}

We have found polarization modes of gravitational waves in
topologically massive and new massive gravities by using the
Newman-Penrose formalism. As was shown in Fig. 1,  the number of
polarization modes is two [$(b)$ and $(c)$]  for the new massive
gravity and one [$(b)$ or $(c)$] for the topologically massive
gravity. In order to obtain  the explicit mode shape, we note that
the linearized equation is equivalent to the FP massive equations
which define the massive spin 2 field in three dimensions. Then,
using the Newman-Penrose formalism, we obtain all polarization modes
of NMG and TMG. As far as we know, this work firstly shows what kind
of massive modes are propagating in the three-dimensional Minkowski
spacetimes.

Some people have considered the TMG and the NMG as toy models of
quantum gravity because they are free from the tachyon and ghosts in
the metric-perturbation approach. Furthermore, it was shown that the
NMG is free from the non-linear ghost (Boulware-Deser ghost) to any
order beyond the decoupling limit~\cite{RGPTY}. It shows really that
the NMG represents a completely consistent ghost free theory of a
fully interacting massive graviton in three dimensions~\cite{Hint}.
It was claimed that this theory is renormalizable~\cite{Oda}.  On
the contrary, this theory seems not be  renormalizable because the
scalar graviton killed by choosing $\alpha=-3\beta/8$ is necessary
to achieve the renormalizability~\cite{stelle}. Actually, the
unitarity and renormalizability exclude to each other~\cite{BHTpro}.
At this stage, we do not have any definite answer to the
renormalizability  of the NMG.

On the other hand, the relevant quantity to specify polarization
modes is not the metric tensor $h_{\mu\nu}$  but the linearized
Ricci tensor $R^L_{\mu\nu}$ (equivalently, the linearized Einstein
tensor $G^L_{\mu\nu}$ with $R^L=0$). If one considers $R^L_{\mu\nu}$
as the physical quantity, the fourth-order linearized Einstein
equation for $h_{\mu\nu}$  reduces to the second-order equation for
$R^L_{\mu\nu}$, which is exactly the self-dual field equation of the
FP massive equation. The latter defines the spin of massive modes.
In this sense, the NMG and TMG are considered as models of truly
massive gravity in three dimensions.   These two theories reflect
how the lower-dimensional gravity could be realized only as massive
gravity theory  because the Einstein gravity is trivial in three
dimensions. The only and nontrivial way to provide the spin 2 field
is to use the linearized Ricci tensor through the Newman-Penrose
formalism. Hence, the TMG  becomes  the first-order theory and  the
NMG is the second-order theory, even though they of TMG and NMG  are
third- and fourth-order theory in the metric-perturbation approach.

 \vspace{1cm}

{\bf Acknowledgments}

This work was supported by the National Research Foundation of Korea
(NRF) grant funded by the Korea government (MEST) through the Center
for Quantum Spacetime (CQUeST) of Sogang University with Grant
No.2005-0049409. Y. Myung  was partly  supported by the National
Research Foundation of Korea (NRF) grant funded by the Korea
government (MEST) (No.2011-0027293).
\newpage
\section*{Appendix: The  linearized perturbation equations in  the TMG}
The explicit forms of linearized equation (\ref{tmgRmunu}) are given
by
\begin{eqnarray}
R_{tt}^{L}&=&\frac{1}{\mu}\left(\partial_xR_{zt}^{L}-\partial_zR_{xt}^{L}\right)\nn\\
R_{tx}^{L}&=&\frac{1}{\mu}\left(\partial_xR_{zx}^{L}-\partial_zR_{xx}^{L}\right)~~{\rm
or}~~
\frac{1}{\mu}\left(\partial_xR_{zt}^{L}-\partial_zR_{xt}^{L}\right)\nn\\
R_{tz}^{L}&=&\frac{1}{\mu}\left(\partial_xR_{zz}^{L}-\partial_zR_{xz}^{L}\right)~~{\rm
or}~~
\frac{1}{\mu}\left(\partial_xR_{tt}^{L}-\partial_tR_{xt}^{L}\right)\nn\\
R_{xx}^{L}&=&\frac{1}{\mu}\left(\partial_tR_{zx}^{L}-\partial_zR_{xt}^{L}\right)\nn\\
R_{xz}^{L}&=&\frac{1}{\mu}\left(\partial_tR_{zz}^{L}-\partial_zR_{tz}^{L}\right)~~{\rm
or}~~
\frac{1}{\mu}\left(\partial_xR_{tx}^{L}-\partial_tR_{xx}^{L}\right)\nn\\
R_{zz}^{L}&=&\frac{1}{\mu}\left(\partial_xR_{tz}^{L}-\partial_tR_{xz}^{L}\right)\nn
\end{eqnarray}
where $\epsilon^{txz}=1$ and the second terms of the r.h.s. in
$R_{tx}^{L},~R_{tz}^{L},$ and $R_{xz}^{L}$ come from the Bianchi
identity  of
$\partial^{\mu}\left(R_{\mu\nu}^{L}-\eta_{\mu\nu}R^{L}/2\right)=0$.

\end{document}